\documentclass[aps,prd,reprint,preprintnumbers,nofootinbib]{revtex4-2}
\usepackage{cancel}

\usepackage{ dsfont }
\usepackage{enumerate}
\usepackage{amsmath}
\usepackage{amsfonts} 
\usepackage{amssymb}
\usepackage{graphicx}
\usepackage{hyperref}
\usepackage{tensor}
\usepackage{siunitx}
\usepackage{float}

\newcommand{\bk}[1]{{\color{black} #1}}

\usepackage{footmisc}

\newcommand{\red}[1]{\textcolor{red}{#1}}

\newcommand{\black}[1]{\textcolor{black}{#1}}

\begin{document}

\title{\Large Dimensionally reducing the classical Regge growth conjecture}
\author{\large Joan Quirant}
\affiliation{Department of Physics, Ben-Gurion University of the Negev, Beer-Sheva 84105, Israel}

\begin{abstract}
\vspace{0.5cm}
We explore the classical Regge growth conjecture in the 4d effective field theory that results from compactifying $D$-dimensional  General Relativity on a compact, Ricci-flat manifold. While the higher dimensional description is given in terms of pure Einstein gravity and the conjecture is automatically satisfied,   it imposes several non-trivial constraints in the 4d spectrum. Namely, there must be either none or an infinite number of massive spin-2 modes, and  the mass ratio between  consecutive Kaluza-Klain spin-2 replicas is bounded by the 4d coupling constants. 
\vspace{1cm}
\end{abstract}

\maketitle
\section{Introduction}

In the recent work \cite{Kundu:2023cof} -see \cite{Chowdhury:2023fwb} for related ideas- it was studied if a gravitational effective field theory (EFT) that includes a massive spin-2 particle in its spectrum was compatible with the   Classical Regge Growth conjecture (CRG)  \cite{Chowdhury:2019kaq}. This conjecture states that the classical (tree-level) S-matrix $\mathcal{A}(s, t)$ of any consistent theory can never grow faster than $s^2$ in the Regge limit, that is, at large $s$ and fixed and physical $t$  -with $s$ and $t$  the usual Mandelstam variables-. In terms of equations, it states
\begin{align}
\label{eq:crgbound}
\lim_{s\rightarrow \infty,\,  t<0 \text{ fixed}} \frac{ \mathcal{A}(s, t)}{s^3} =0\, ,
\end{align}
where by $s\rightarrow \infty$ we mean  $\Lambda\gg s\gg |t|$, with $\Lambda$  the cut-off of the EFT considered.

The main conclusion of \cite{Kundu:2023cof} was to show the incompatibility of the setup described above with the CRG conjecture. If the CRG holds, a gravitational EFT which includes a massive spin-2 particle -and no other higher spin particles- would be in the swampland \cite{Vafa:2005ui}.\footnote{This conclusion is in tension with \cite{Chowdhury:2023fwb}, where a particular choice of constants for de Rham-Gabadadze-Tolley  (dRGT) massive gravity was argued to be consistent with the CRG conjecture. This discrepancy, which deserves further clarification, does not affect the results of this paper, since they can be derived independently of \cite{Kundu:2023cof} and \cite{Chowdhury:2023fwb}.}

This result was in line with the so-called  spin-2 swampland conjecture \cite{Klaewer:2018yxi, DeRham:2018bgz,Palti:2019pca} and with the recent works studying the (in)consistency of massive gravity -see \cite{Bonifacio:2016wcb,deRham:2017xox,Bellazzini:2017fep,deRham:2018qqo,Alberte:2019xfh,Wang:2020xlt, Bellazzini:2023nqj, Chowdhury:2023fwb} for a biased selection and \cite{Hinterbichler:2011tt, deRham:2014zqa} for reviews-.

Regarding the state of the CRG conjecture, though a complete demonstration is still lacking, there is strong evidence in favour of it. Using the duality between Anti-de
Sitter space (AdS) and conformal field theories, it has been proven in \cite{Chandorkar:2021viw} that, in the dual picture, the CRG conjecture follows  from  the chaos bound of \cite{Maldacena:2015waa}. In flat space, it was shown in \cite{Haring:2022cyf}  that  the scattering of scalar particles in dimensions bigger or equal to five satisfies it. Here, as we did in \cite{Kundu:2023cof}, we will limit ourselves to assuming its validity, studying the consequences that derive from it.

This being said, the logical next step after \cite{Kundu:2023cof} is to consider an EFT with not only one but any number of massive spin-2 particles in the spectrum. This is a common ingredient in theories with extra dimensions, where in the 4d EFT the graviton comes typically accompanied by an infinite tower of massive spin-2 particles, its  Kaluza-Klain  (KK) replicas.\footnote{When the length of the internal $l_{\rm internal}$ and external $l_{\rm external}$ dimensions satisfy $l_{\rm external}\gg l_{\rm internal}$, the mass of the Kaluza-Klain spin-2 replicas usually becomes much bigger than the energy scale probed in the EFT and the massive spin-2 states can be ignored in the low-energy description.} In this paper, we would like to understand how  the CRG conjecture is satisfied in these scenarios.%, as we expect it to be.  

To do so, we will focus on a very concrete but general model:  we will study General Relativity (GR) dimensionally reduced to four dimensions. Of course, GR in $\mathds{R}^{1,D-1}$ trivially satisfies the CRG conjecture, the $2\rightarrow 2$ scattering of a GR graviton scales with $s$ in the Regge limit at most as $\mathcal{A}\sim s^2$. The point is that when GR is compactified to 4d (we go from  $\mathds{R}^{1,D-1}$ to $\mathds{R}^{1,3}\times X_{D-4}$), the description is given not only in terms of a graviton but it includes an infinite tower of massive spin-2 particles. %Since we are just changing the vacuum where the theory is being studied, we expect the bound \eqref{eq:crgbound} to continue to be true for any amplitude. 
This provides an arena where  the CRG conjecture can be \emph{tested} in the presence of several massive spin-2 states.

What we will see in this work is that the CRG conjecture imposes non-trivial constraints\footnote{These constraints will be automatically satisfied once a valid internal geometry is specified.} on the particle content of the 4d description. They teach us how the CRG requirement can be fulfilled in a  4d EFT containing massive spin-2 particles. Namely, as we will see in section \ref{sec:gr}, in the spectrum of the 4d effective field theory:
\begin{itemize}
\item There must be either none or an infinite number of massive spin-2 particles. This is in line with the absence of consistent finite truncations of the graviton tower, already discussed in the literature \cite{Duff:1989ea, Gauntlett:2009zw, Liu:2010ya, Bonifacio:2019ioc}.
\item The  mass ratio between consecutive massive spin-2 replicas is  bounded by the 4d couplings constants. Similar results imposing unitarity  were derived in \cite{Csaki:2003dt, Bonifacio:2019ioc}.
\end{itemize}

Before presenting these results, we will start by the beginning, briefly recalling the work done in \cite{Kundu:2023cof}.

\section{A single massive spin-2 particle}
\label{sec:recap}

In \cite{Kundu:2023cof} it was studied the tree-level $2\rightarrow2$ scattering of a massive spin-2 particle in a theory containing neither other massive spin-2 states nor higher-spin particles. We wanted to check if this setup was compatible with the CRG conjecture. To do so:
\begin{itemize}
\item We assumed that the spin-2 particle could couple to a graviton, a (massive or massless) scalar particle, and a massive spin-1 particle.\footnote{Symmetries forbid interactions between two identical massive spin-2 particles and one massless spin-1 field or one fermion.}
\item We considered both parity-even and parity-odd interactions.
\item We included all contact terms with an arbitrary but finite number of derivatives.
\end{itemize}
Exchange diagrams and contact terms are the two sources of contributions to any classical two-to-two scattering amplitude. Both can be computed directly using on-shell methods, in a Lagrangian independent way, as explained in \cite{Costa:2011mg}.
\begin{itemize}
\item Exchange diagrams can be built from the on-shell cubic couplings. First, one has to list all possible on-shell three-point interactions between two massive spin-2 particles and the exchanged particle. Then, two sets of these vertices (multiplied by arbitrary constants) are connected through the correspondent propagator. In four dimensions, we found 24 independent exchange pieces, reproducing the results of \cite{Hinterbichler:2017qyt, Bonifacio:2017iry, Bonifacio:2018vzv, Bonifacio:2018aon, Bonifacio:2019mgk}.
\item Contact terms are a bit more tricky since, in principle, one can construct infinitely many of them, introducing more and more derivatives. In \cite{Kundu:2023cof}, adapting the ideas developed in \cite{Bonifacio:2018vzv}, we included all contact interactions with  an arbitrary, but finite, number of derivatives.
\end{itemize} 
With all these ingredients, we showed in \cite{Kundu:2023cof} that a gravitational theory of a single massive spin-2 particle, coupled to any other state of spin $< 2$, can never be made consistent with the CRG conjecture.

\section{Several massive spin-2 particles}
\label{sec:several}

We will now explain how to generalise the results of the previous section to incorporate  any number of massive spin-2 particles in the spectrum. As we will see, the modifications are conceptually quite simple but technically very involved.

\begin{figure}[h!]
\centering
\includegraphics[width=6cm]{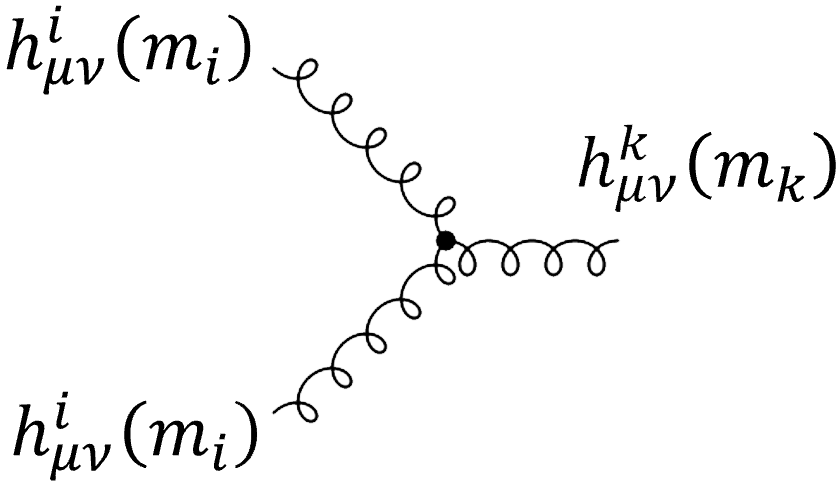}
\caption{Cubic interaction between two identical ($h^i_{\mu\nu}$ with mass $m_i$), one different ($h^k_{\mu\nu}$ with mass $m_k\neq m$) massive spin-2 particles.}
\label{figure:cubichhh}
\end{figure}

The \emph{only} novelty concerning the previous computation lies in the number of the allowed cubic couplings. Besides the 24 previous pieces,  we must include interactions between two identical and one  different massive spin-2 particles, as shown in figure \ref{figure:cubichhh}.

In four dimensions there are  seven parity-even and nine parity-odd independent on-shell three-point functions of this kind, some of them already computed in \cite{Bonifacio:2019ioc}. We give the complete list with the explicit expressions in appendix \ref{app:interactions}, \bk{where we also derive a Lagrangian basis for the party-even contributions}. For practical proposes, let us denote this set of interactions by
\begin{align}
\label{eq:twomassive}
\mathcal{A}\left( h^i(m_i), h^i(m_i), h^k(m_k)\right)\equiv \sum_{j=1}^{16} c^{ii}_{k,j} f_j\left(h^i, h^i, h^k\right)\, ,
\end{align}
where the $ c^{ii}_{k,j}$ are arbitrary constants and the $f_j\left(h^i, h^i, h^k\right)$ are the ($9$ parity-even and $7$ parity-odd) on-shell cubic amplitudes, given in appendix \ref{app:interactions}.

This is the first step in accounting for several massive spin-2 particles, but it is not the end of the story. What we have just described corresponds to a theory in which any pair of identical massive spin-2 particles $\{h^i_{\mu\nu},h^i_{\mu\nu}\}$ only interacts with  one different massive spin-2 state $\{h^k_{\mu\nu}\}$. If we want to include the possibility that they couple to  any number of different massive spin-2 particles, we need to replace \eqref{eq:twomassive} with 
\begin{align}
\label{eq:severalmassive}
\sum_{k=1}^{K^{ii}}\mathcal{A}\left( h^i(m_i), h^i(m_i), h^k(m_k)\right)=\sum_{k=1}^{K^{ii}}\sum_{j=1}^{16} c^{ii}_{k,j} f_j\left(h^i, h^i, h^k\right)\, ,
\end{align}
where the index $k=1,\dots K^{ii}$ describes the coupling with $K^{ii}$ distinguishable ($m_i\neq m_k$) massive spin-2 particles.

Generalising \cite{Kundu:2023cof} to include any number of massive spin-2 fields would correspond to take $K^{ii}=\rm{arbitrary}$ and $c^{ii}_{k,j}=\rm{arbitrary}$, since this is the most general possibility. Unfortunately, this case is technically very complicated, and little can be done explicitly. It would require introducing an arbitrarily large number of new constants in the equations of \cite{Kundu:2023cof}, which were already very complex.

To understand whether the CRG conjecture can be satisfied in the presence of several massive spin-2 particles, we find starting with a simpler model more illuminating. As a proof of concept example, we will study the 4d effective field theory obtained after dimensionally reducing GR. In this case, the $c^{ii}_{k,j}$ are not arbitrary: they are completely fixed  once the internal manifold is specified -GR has no free parameters-and there will be relations among them. Similar ideas studying the unitarity of GR under dimensional reductions were derived in \cite{Bonifacio:2019ioc}, which we will use in this note.

\section{Proof of concept: general relativity}
\label{sec:gr}

We will start by commenting and motivating again why this example is interesting. The framework described here is a summary of \cite{Hinterbichler:2013kwa, Bonifacio:2019ioc}, which we refer the reader for a more detailed discussion  -we will only introduce  the minimal ingredients to make the note  self-contained-. 

Consider the Einstein-Hilbert action in $D>4$ dimensions
\begin{align}
\label{eq:einsteinacion}
\mathcal{L}=\frac{M_D^{D-2}}{2}\sqrt{-G}R(G)\, ,
\end{align}
with $M_D$ the $D$-dimensional Plank mass. In $\mathds{R}^{1,D-1}$ this theory is, of course, consistent with the CRG conjecture \cite{Chowdhury:2019kaq}. Dimensionally reducing it to 4d while keeping all massive modes  just means selecting a different background for the theory. Consequently, one would expect the CRG conjecture to continue to be satisfied in the 4d picture. The interesting point is that, while in $D$ dimensions we have a description in terms of  pure (Einstein) gravity, in 4d the dimensional reduction of the graviton \emph{produces} a graviton but also a tower of massive spin-2, spin-1 and scalar particles. We can then take any Kaluza-Klain  massive spin-2 copies of the graviton and compute its $2\rightarrow 2$ scattering. In a generic theory with no other massive spin-2 particles, we saw in \cite{Kundu:2023cof} that this scattering would violate the CRG bounds. In contrast, we will see below how the CRG conjecture is satisfied in this set-up, imposing restrictions in the \emph{effective} spectrum.

\subsection{Dimensionally reduced theory}
\label{subsec:dimrethe}
We will study the Lagrangian \eqref{eq:einsteinacion} in the direct product space $\mathcal{M}_D=\mathds{R}^{1,3}\times X_{D-4}$, with metric
\begin{align}
ds^2=G_{AB}dX^A dX^B=\eta_{\mu\nu} dx^\mu dx^\nu+g_{ab}dy^ady^b\, ,
\end{align}
where $A=1,\dots, D$, $\mu=1,\dots 4$, $a=D-4,\dots, D$ and we  require $X_{D-4}$ to be a closed, smooth, connected, orientable Ricci-flat\footnote{This is necessary to solve the vacuum equations.} Riemannian manifold. To obtain the interactions in the lower dimensional description, one first needs to expand the metric around the background $\bar{G}_{AB}$
\begin{align}
G_{AB}=\bar{G}_{AB}+\frac{2}{M_D^{\frac{D-2}{2}}}\delta G_{AB}\, ,
\end{align}
and then expand the fluctuations using the usual Hodge decomposition. Skipping some field redefinitions and showing only the contributions relevant to our computations, we have
\begin{subequations}
\label{eq:expansion}
\begin{align}
\label{eq:expm2}
\delta G_{\mu\nu}(x,y)&=\sum_n h^n_{\mu\nu}(x)\psi_n(y)+\frac{1}{\sqrt{V}}h^0_{\mu\nu}(x)\, ,\\
\delta G_{\mu a}(x,y)&=\sum_i A^i_\mu(x) Y_{a,i}(y)+\dots\, \\
\delta G_{ab}(x,y)&=\frac{1}{D-4}\frac{1}{\sqrt{V}}\phi^0(x)g_{ab} +\dots\, ,
\end{align}
\end{subequations}
where 
\begin{equation}
V=\int_{ X_{D-4}} \sqrt{-g}dy^a\equiv \int_{ X_{D-4}} d\text{vol}_{X_{D-4}}
\end{equation}
and the  $\{\psi_n,\, Y_{n,i}\}$ satisfy
\begin{subequations}
\begin{align}
\Delta\psi_n\equiv-\square \psi_i &=m_n^2\psi_i\, ,  \\[1.1ex]
\int_{ X_{D-4}}\psi_n\psi_m\, d\text{vol}_{X_{D-4}}&=\delta_{nm}\, , \quad m_n^2>0\, ;\\[1.1ex]
\Delta Y_{a,i}\equiv-\square Y_{a,i}+R_a^bY_{b,i} &= m_i^2Y_{a,i}\, ,	\\[1.1ex]
\int_{ X_{D-4}}Y^a_{i}Y_{a,j}\, d\text{vol}_{X_{D-4}}&=\delta_{ij}\, , \quad m_i^2\geq 0\, ,
\end{align}
\end{subequations}
being $R_{ab}$ the internal Ricci curvature. We refer again to \cite{Hinterbichler:2013kwa, Bonifacio:2019ioc} for a more detailed discussion of all the quantities and definitions. Plugging all these expressions into \eqref{eq:einsteinacion} and expanding the action, one can obtain the spectrum and the interactions of the dimensionally reduced theory. Let us summarise the main results we will need.

\subsubsection{Spectrum}

From the quadratic terms one can see that the four-dimensional theory contains:
\begin{itemize}
\item One massless graviton, $h^0_{\mu\nu}$.
\item A tower of massive spin-2 particles $h^n_{\mu\nu}$ with squared masses $m_n^2>0$. They come from the eigenfunctions of the scalar Laplacian on $X_{D-4}$.
\item A tower of spin-1 fields $A^i_\mu$ with squared masses $m_i^2\geq 0$. They come from the eigenfunctions of the vector Laplacian on $X_{D-4}$. This tower includes the killing vectors, which are massless.
\item A massless scalar field $\phi^0$ controlling the internal volume.
\end{itemize}
There are more scalar fields in the spectrum coming from the terms omitted in the decomposition of $\delta G_{ab}$. We will ignore them since they will not play any role in our computation -see appendix \ref{app:cubic} for the details-.

Finally, let us remember that the relation between the higher dimensional ($M_D$) and the lower dimensional ($M_d$) Planck mass is given by
\begin{align}
M_d^{d-2}=VM_D^{D-2}\, .
\end{align}

\subsubsection{Cubic interactions}
\label{subsec:cubicint}
\begin{figure}[h!]
\centering
\includegraphics[width=8cm]{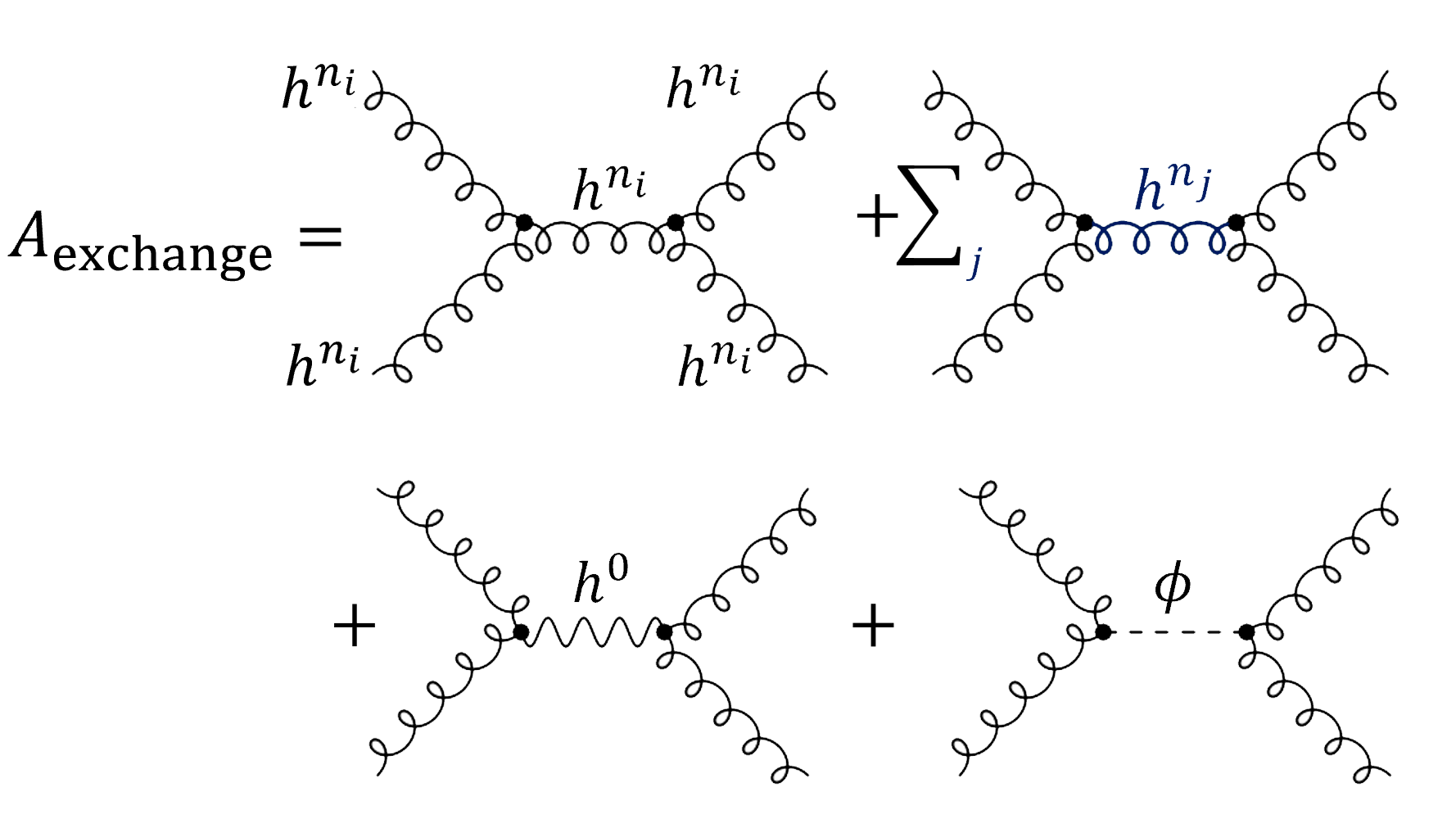}
\caption{Exchange contributions to the $2\rightarrow2$ scattering of a massive particle $h^{n_i}$. The particle is a  part of the KK tower of the graviton in 4d. In the picture $n_j\neq n_i$.}
\label{figure:exchange}
\end{figure}

We are  interested in the (classical) scattering  $h^{n_i}h^{n_j}\rightarrow h^{n_k}h^{n_l}$. Therefore,  we will only need the three-point functions involving two (on-shell) massive spin-2 fields to construct the exchange diagrams. We relegate the explicit expressions to appendix \ref{app:cubic}, while we list here the  relevant interactions:
\begin{itemize}
\item Three massive spin-2 particles: $\mathcal{A}\left( h^{n_1}, h^{n_2}, h^{n_3}\right)$.
\item Two identical  massive spin-2 particles -otherwise this interaction vanishes, see \eqref{eq:hhho}-  and the graviton: $\mathcal{A}\left( h^{n_1}, h^{n_1}, h^{0}\right)$.
\item Two distinct massive spin-2 particles -otherwise this interaction vanishes, see \eqref{eq:hha}-  and a spin-1 particle: $\mathcal{A}\left( h^{n_1}, h^{n_2}, A^i_\mu\right)$.
\item Two massive spin-2 particles and a scalar field: $\mathcal{A}\left( h^{n_1}, h^{n_2}, \phi \right)$
\end{itemize}
A pictorial representation of the exchange contributions to the $h^{n_i}h^{n_i}\rightarrow h^{n_i}h^{n_i}$ scattering  can be seen in figure \ref{figure:exchange}.
It is also worthy to define in this section the triple overlap integrals $g_{n_1n_2n_3}$
\begin{align}
\label{eq:triple}
g_{n_1n_2n_3}=\int_{ X_{D-4}}\psi_{n_1}\psi_{n_2}\psi_{n_3}\, d\text{vol}_{X_{D-4}}
\end{align}
-the $\psi_{n_1}$ introduced in expression \eqref{eq:expm2} - which will be used later on.

\subsubsection{Contact terms}

Finally, to compute any $h^{n_i}h^{n_j}\rightarrow h^{n_k}h^{n_l}$ scattering,  we need the (on-shell) 4-point interactions. Repeating the previous game, one has to insert the decompositions introduced in section \ref{subsec:dimrethe} into the action \eqref{eq:einsteinacion} and collect the terms involving four massive spin-2 particles. We write the explicit form of this interaction, which we denote by $\mathcal{A}_{\text{contact}}\left( h^{n_1}, h^{n_2}, h^{n_3}, h^{n_4}\right)$,   in appendix \ref{app:contact}. For posterior uses, we  define here the quartic overlap integrals
\begin{align}
\label{eq:quartic}
g_{n_1n_2n_3n_4}=\int_{ X_{D-4}}\psi_{n_1}\psi_{n_2}\psi_{n_3}\psi_{n_4}\, d\text{vol}_{X_{D-4}}\, ,
\end{align}
which, as discussed in \cite{Bonifacio:2019ioc},  can be written in terms of the cubic overlap integrals
\begin{align}
\label{eq:quarticcubic}
&g_{n_1n_2n_3n_4}=\sum_i g_{n_1n_2n_i}g_{n_3n_4n_i}+\frac{1}{V}\delta(n_1n_2)\delta(n_3n_4)\nonumber\\&=\sum_i g_{n_1n_3n_i}g_{n_2n_4n_i}+\frac{1}{V}\delta(n_1n_3)\delta(n_2n_4)\nonumber\\&=\sum_i g_{n_1n_4n_i}g_{n_2n_3n_i}+\frac{1}{V}\delta(n_1n_4)\delta(n_2n_3)\, .
\end{align}

\subsection{Results}
\label{subsec:resuls}
Having introduced all the necessary ingredients, we are finally in the position to test the CRG conjecture in the EFT obtained from the dimensional reduction of   GR. Let us briefly recall what are the  steps to follow:
\begin{enumerate}
\item Compute the  tree level $h^{n_i}h^{n_j}\rightarrow h^{n_k}h^{n_l}$ scattering. Using the language presented in the previous section, it  reads
\begin{align}
\mathcal{A}\left( h^{n_i}, h^{n_j}, h^{n_k}, h^{n_l}\right)&=\mathcal{A}_{\text{contact}}\nonumber+\mathcal{A}_{\rm exhange}\, ,
\end{align}
where, to construct the exchange diagrams, we take two sets of the three-point functions introduced in section \ref{subsec:cubicint}, ``remove" the exchanged leg, and connect them through the correspondent propagator. 
\item Expand the total amplitude in the limit $s\gg t$, where $s$ and $t$ are the usual Mandelstam variables
\begin{align}
&\lim_{s\gg t}\mathcal{A}\left( h^{n_i}, h^{n_j}, h^{n_k}, h^{n_l}\right)=\mathcal{A}_0(t)s^0+\mathcal{A}_1(t)s^1\nonumber\\&+\mathcal{A}_2(t)s^2+\mathcal{A}_3(t)s^3+\dots\, .
\end{align}
In appendix \ref{app:kinematics} we recall the  definition of the Mandelstam variables and set the conventions for the kinematics.
\item Finally, from the previous expansion we impose that 
\begin{align}
\label{eq:constraintt}
\mathcal{A}_n(t)=0\, , \{n\geq 3,\, \forall t\}\, .
\end{align}
\end{enumerate}
We will do this for any of the $5^4=625$ choices of polarisation of the scattered spin-2 particles.\footnote{Not all choices are independent, some of them will be related by crossing symmetry.} 

Taking $n_i=n_j=n_k=n_l$ equation \eqref{eq:constraintt} requires
\begin{align}
\label{cons1}
\forall n_i\, , \, 4m_{n_i}^2g_{n_in_in_in_i}-\sum _k 3m_{n_k}^2g_{n_in_in_k}^2=0\, ,
\end{align}
which can also be written, using expansion \eqref{eq:quarticcubic}, as
\begin{align}
\label{cons2}
\forall n_i\, , \, \sum _k \left(4m_{n_i}^2-3m_{n_k}^2\right)g^2_{n_in_in_k}+4m_{n_i}^2 V^{-1}=0\, .
\end{align}
These relations, which will be automatically satisfied by any valid internal geometry, have several consequences regarding the 4d spectrum. They are not completely  new since they also appear when one demands the dimensional reduced theory to be unitary  \cite{Csaki:2003dt, Bonifacio:2019ioc} -actually, in \cite{Bonifacio:2019ioc} they were even able to find stronger conditions-.\footnote{This is because, while the CRG conjecture \emph{cares} about terms scaling with the energy at order  $s^3\sim E^6$ or higher, unitarity in this context places conditions on terms scaling with the energy at order $E^4$ or higher.}

From \eqref{cons2} it can be deduced that there must be an infinite number of KK modes in the spectrum. Since the term outside the sum is positive definite, the sum itself must produce a negative contribution that compensates it. This implies that for any $h^{n_i}$   there must exist some $h^{n_l}$  to which the $ h^{n_i}$ couples (that is, $g_{n_in_in_l}\neq 0$) and such that 
\begin{align}
\label{eq:masscons}
\frac{2}{\sqrt{3}} m_{n_i}<m_{n_l}\, .
\end{align}
Taking $m_{n_i}=m_{n_1}$, this equation tells us that there must exist another spin-2 particle with mass  $m_{n_2}>m_{n_1}$ in the spectrum. We can then apply the same strategy to $m_{n_2}$ and conclude that the spectrum must contain a third spin-2 state with mass $m_{n_3}>m_{n_2}$. Repeating this reasoning, we see that any finite truncation of the  KK tower of the graviton is incompatible with the CRG  conjecture. This goes in the lines of  \cite{Duff:1989ea} -see also \cite{, Gauntlett:2009zw, Liu:2010ya}- who first showed the inconsistencies of a truncated KK spin-2 spectrum  by using the breaking of the massive gauge invariances.

On the other side,  equation  \eqref{cons1} is useful to see\footnote{This  can also be seen from \eqref{cons2} since they are equivalent, eq. \eqref{cons1} just gives a cleaner expression.} that the mass ratio of consecutive spin-2 modes is bounded by the 4d couplings. Since the first term in \eqref{cons1} is positive definite, the second term cannot be ``too negative''. In other words, $\forall h^{n_i}$ there must exist some  $h^{n_k}$ to which the  $h^{n_i}$ couples (that is, $g_{n_in_in_k}\neq 0$)  and such that 
\begin{align}
\label{eq:forbgap}
\frac{m_{n_k}^2}{m_{n_i}^2}\leq \frac{4}{3}\frac{g_{n_in_in_in_i} }{g_{n_in_in_k}^2} >1\, ,
\end{align}
constraining the mass ratios of consecutive spin-2 KK particles.\footnote{If $m_{n_k}=m_{n_{i+1}}$ then this is the maximum allowed gap between consecutive massive spin-2 states. If not, this means that $m_{n_k}>m_{n_{i+1}}$ -since $m_{n_k}>m_{n_{i}}$-  and so the gap is even smaller.}

A consequence of the  relation \eqref{eq:forbgap} is that it seems difficult to generate a consistent gravitational  4d theory in which part of the graviton KK  tower can be integrated out leaving a finite -bigger than zero- number of massive spin-2 particles in the spectrum. For this to make sense, the mass $m_\Lambda$ of the lightest integrated particle should satisfy   $m, E << \Lambda_{\rm{EFT}}<< m_\Lambda$ where $m$ is the mass scale of the particles kept in the theory, $E$ is the energy at which the theory is being probed and $\Lambda_{\rm{EFT}}$ is the cut-off of the EFT. What we learn from \eqref{eq:forbgap} is that the 4d coupling constants bound the gap between the mass of the  spin-2 replicas, so the couplings should be appropriately tuned to achieve the desired \emph{mass  separation}. Once chosen, one should find the concrete geometry producing these values for the 4d couplings, which can be a very non-trivial task.

Equation \eqref{cons2} -or its equivalent expression \eqref{cons1}- and its consequences are the main result of this paper. They teach us that, even if we start with a theory compatible with the CRG conjecture, as it is GR, when it is dimensionally reduced to 4d the spectrum of the resulting theory satisfies two non-trivial constraints:
\begin{enumerate}
\item Either there is none or an infinite number of massive spin-2 modes.
\item The gap in the mass ratio of consecutive KK spin-2 states is bounded by the coupling constants of the theory.  
\end{enumerate}

We derived all these conditions by looking at the $h^{n_i}h^{n_i}\rightarrow h^{n_i}h^{n_i}$ amplitude, so one could wonder about the  more general $h^{n_i}h^{n_j}\rightarrow h^{n_k}h^{n_l}$ interaction. We also studied the CRG conjecture  for this case. Nevertheless, the results and constraints derived from  it are less powerful and interesting than the ones we already presented. In any case, the reader interested can find an ancillary \texttt{Mathematica} notebook with the code used.\footnote{\black{As a double-check of this code, we verified that in the high energy limit - that is in the $\{s\rightarrow\infty\, ,\, t\rightarrow \infty\}$ limit- it reproduces the results of \cite{Bonifacio:2019ioc}.}}

Before moving to the conclusions, it is worth pausing here for a moment to make a couple of comments.

As pointed out throughout the section, the constraints  \eqref{cons1}-\eqref{cons2}, imposed by the CRG conjecture, had already appeared in the literature. They are also a requirement for the dimensionally reduced theory to be unitary \cite{Csaki:2003dt, Bonifacio:2019ioc} -which actually demands more stringent conditions-. On the other hand, a condition similar to \eqref{eq:masscons} was derived in \cite{Duff:1989ea} by studying the gauge invariances of a dimensionally reduced theory when the KK tower of the graviton is truncated. The novelty here is that we have derived all these requisites using the CRG conjecture. This is  a non-trivial check of the conjecture: for the first time it has been tested under dimensional reduction.

We have studied  the dimensional reduction of  GR on a compact, Riemannian, Ricci flat internal manifold down to a 4d flat space. A natural question is thus how the conclusions would change if we modified any of the ingredients: including  matter  or higher-derivative corrections, choosing a different external space... These considerations can be taken into account all at once by studying the most generic case,  discussed in section \ref{sec:several},  which is a formidable task. A more doable approach could be to study the changes one by one, for instance by looking at the scattering of massive spin-2 particles in AdS or by starting from GR coupled to some matter. Based on the results of \cite{Kundu:2023cof} and on the apparent impossibility of constructing truncations with a finite number of massive spin-2 modes \cite{Duff:1989ea, Liu:2010ya}, we would expect the conclusions obtained here to hold in more general scenarios. We leave the exploration of these ideas for future work.

\section{Conclusions}

In this note, we have studied the CRG conjecture in the 4d effective field theory that results from compactifying  $D$-dimensional General Relativity (with $D>4$) on a closed, Ricci-flat manifold. To do so, we have used the tools and the framework developed in \cite{Bonifacio:2019ioc}.

Whereas the conjecture is trivially satisfied in the $D$-dimensional description\footnote{The $2\rightarrow 2$ scattering of a GR graviton in $\mathds{R}^{1,D-1}$ scales with $s$ in the Regge limit as $\mathcal{A}\sim s^n\,,\, n\leq 2$, \cite{Chowdhury:2019kaq}.}, the 4d picture consists of a theory of gravity coupled to an infinite number of  massive spin-2 particles. We already saw in \cite{Kundu:2023cof} -see \cite{Chowdhury:2023fwb} for related work- that any gravitational EFT containing a single massive spin-2 particle cannot be made consistent with the CRG conjecture. The example studied here, in contrast, serves as an arena to see how the CRG conjecture is satisfied in the  presence of several massive spin-2 states.

The main result of this work is equation \eqref{cons1} -or equivalently equation \eqref{cons2}-  which is required for the CRG conjecture to hold in the 4d framework. Both conditions are automatically met when choosing a valid internal geometry. From the  4d perspective, they teach us how the CRG conjecture  can be realized  in a theory containing  massive spin-2 particles. These expressions are also part of the conditions for the theory to be unitary  \cite{Csaki:2003dt, Bonifacio:2019ioc}. Two consequences follow from them:
\begin{itemize}
\item The 4d spectrum must include either no massive spin-2 fields or an infinite number of them, a  finite truncation would not be possible. This was also discussed from other points of view in \cite{Duff:1989ea, Gauntlett:2009zw, Liu:2010ya, Bonifacio:2019ioc}.
\item The mass ratio between the consecutive KK spin-2 replicas is bounded by the 4d coupling constants.  We concrete this point in section \ref{subsec:resuls}. Similar conclusions (actually a   bit stronger)  were derived in \cite{Bonifacio:2019ioc} by studying unitarity in the 4d theory.
\end{itemize}
We see, then, that even if we start with a theory satisfying the CRG conjecture in $D$ dimensions, the conjecture imposes non-trivial conditions in the 4d spectrum. This shows the power of the CRG conjecture to discern between consistent 4d theories, in this case regarding the ones with  a higher-dimensional embedding, in line with the spirit of the swampland program \cite{Vafa:2005ui} -see \cite{Palti:2019pca, vanBeest:2021lhn} for reviews-. 

It is important to keep in mind that in this work we have focused on the concrete example of GR dimensionally reduced to a flat 4d background. Therefore, one could wonder about other possibilities: starting with GR plus some matter, including higher-derivative corrections, changing the external space... We explained in section \ref{sec:several} how to address the most generic situation, which would simultaneously encode all these possibilities. Unfortunately, this seems to be a highly complex task, so it may be smarter to add more ingredients one by one. These are exciting scenarios that for sure deserve further investigation.

In any case, we actually expect the conclusions presented here to hold  in more general contexts. In light of the results of \cite{Kundu:2023cof} together with this work, it seems pretty unlikely that the CRG conjecture could be satisfied in a gravitational EFT with a finite number of massive spin-2 particles, at least in flat space. We leave the exploration of these avenues and any other potential cases of interest for  future work.

\vspace{10px}
{\bf Acknowledgements}

We would like to thank Eran Palti for very useful discussions, collaboration and comments on the manuscript. This work is  supported by the Israel Science Foundation (grant No. 741/20) and by the German Research Foundation through a German-Israeli Project Cooperation (DIP) grant “Holography and the Swampland”.

\appendix
\section{Cubic vertices}
\label{app:interactions}

In this appendix, we will discuss the possible three-point amplitudes for three massive spin-2 particles, two of which are identical and different from the third one. We divide this appendix into two sections. In the first  part, \ref{sec:onshell},  we list all the parity-even and parity-odd on-shell three-point interactions. In the second part, \ref{sec:lagrangian}, we give a Lagrangian basis for the parity-even terms.

\subsection{On-shell amplitudes}
\label{sec:onshell}

Here we list all the possible (parity-even and parity-odd) on-shell three-point functions between two identical and one different massive spin-2 particles. In the particular case $d=4$, we also discuss the dimensionally dependent relations, which come from the fact that any set of five or more vectors is linearly dependent in four dimensions. 

Notation: we denote by  $\mathcal{M}_{i,j,k}\left(m_i,m_j,m_k\right)$ the on-shell three-point amplitude involving three particles of spin $\{i, j, k\}$, mass $\{m_i, m_j, m_k\}$, polarisation matrices $\{\epsilon_1, \epsilon_2, \epsilon_3 \}$ and momentum $\{p_1, p_2, p_3 \}$. We define $A_{ij}\equiv \epsilon_i\cdot p_j$ and $B_{ij}=\epsilon_i\cdot\epsilon_j$.  $\varepsilon$ is the Levi-Civita tensor, $\varepsilon\left(p_i,p_j,\epsilon_k,\epsilon_l\right)\equiv \varepsilon^{\mu\nu\alpha\beta}p_{i\mu}p_{j\nu}\epsilon_{k\alpha}\epsilon_{l\beta}$

\begin{center}
\textbf{Parity even}
\end{center}

There are in general $8$ different parity-even on-shell cubic amplitudes, listed in table \ref{MM2M2}. Part of this classification was already discussed in \cite{Bonifacio:2019ioc}.

\begin{table}[h!]
\centering
\def\arraystretch{1.5}%  1 is the default, change whatever you need
\begin{tabular}{ c}
$\mathcal{M}^{\rm even}_{2,2,2}\left(m,m,m_k\right)$    \\ \hline
$B_{12}B_{23}B_{13}=\mathcal{X}_1$\\
$B_{12}^2A_{31}^2=\mathcal{X}_2$\\
$B_{13}^2A_{23}^2+B_{23}^2A_{12}^2=\mathcal{X}_3$\\
$B_{13}B_{23}A_{12}A_{23}=\mathcal{X}_4$\\
$B_{12}B_{23}A_{12}A_{31}+B_{12}B_{13}A_{23}A_{31}=\mathcal{X}_5$\\
$B_{12}A_{12}A_{21}A_{31}^2=\mathcal{X}_6$\\
$B_{23}A_{12}^2A_{21}A_{31}-B_{13}A_{12}A_{21}^2A_{31}=\mathcal{X}_7$\\
$A_{12}^2A_{21}^2A_{31}^2=\mathcal{X}_8$\\
	\end{tabular}
	\caption{All possible parity even on-shell cubic amplitudes for two identical, one different  massive spin-2 particles.} \label{MM2M2}
\end{table}
In $d=4$ the Gram matrix of the vectors $\{p_i,\epsilon_{i\mu}\}$ must vanish, from which we obtain 
\begin{align}
&(-4 m^2 m_k^2+m_k^4) \mathcal{X}_1+2 m_k^2 \mathcal{X}_2+2 m^2 \mathcal{X}_3\nonumber\\&+(2 m_k^2-4 m^2) \mathcal{X}_4+2 m_k^2
   \mathcal{X}_5+4 \mathcal{X}_6+4 \mathcal{X}_7=0\, ,
\end{align}
which can be used to ignore, for instance, $\mathcal{X}_7$.

\begin{center}
\textbf{Parity Odd}
\end{center}
Regarding the parity odd terms, in general there are 13 distinct possibilities, enumerated in table \ref{MM2M2odd}.
\begin{table}[H]
\centering
\def\arraystretch{1.5}%  1 is the default, change whatever you need
\begin{tabular}{ c}
$\mathcal{M}_{2,2,2}^{\rm odd}\left(m,m,m_k\right)$    \\ \hline
$B_{13} B_{23} \,\varepsilon\left(p_1,p_2, \epsilon_1,\epsilon_2\right)=\tilde{\mathcal{X}}_{1}$	\\
$B_{12} B_{23} \,\varepsilon\left(p_1,p_2,\epsilon_1,\epsilon_3\right)-B_{12} B_{13} \,\varepsilon\left(p_1,p_2,\epsilon_2,\epsilon_3\right)=\tilde{\mathcal{X}}_{2}$	\\
$A_{31} B_{12}\, \left(\varepsilon\left(p_1,\epsilon_1,\epsilon_2,\epsilon_3\right)+\varepsilon\left(p_2,\epsilon_1,\epsilon_2,\epsilon_3\right)\right)=\tilde{\mathcal{X}}_{3}$\\
$A_{21} B_{13} \,\varepsilon\left(p_1,\epsilon_1,\epsilon_2,\epsilon_3\right)-A_{12} B_{23} \,\varepsilon\left(p_2,\epsilon_1,\epsilon_2,\epsilon_3\right)=\tilde{\mathcal{X}}_{4}$\\
$A_{21} B_{13} \,\varepsilon\left(p_2,\epsilon_1,\epsilon_2,\epsilon_3\right)-A_{12} B_{23} \,\varepsilon\left(p_1,\epsilon_1,\epsilon_2,\epsilon_3\right)=\tilde{\mathcal{X}}_{5}$\\
$A_{31}^2 B_{12} \,\varepsilon\left(p_1,p_2, \epsilon_1,\epsilon_2\right)=\tilde{\mathcal{X}}_{6}$\\
$\left(A_{21} A_{31} B_{13} -A_{12} A_{31} B_{23}\right) \,\varepsilon\left(p_1,p_2, \epsilon_1,\epsilon_2\right)=\tilde{\mathcal{X}}_{7}$\\
$A_{31} B_{12}\left(A_{21}  \,\varepsilon\left(p_1,p_2,\epsilon_1,\epsilon_3\right)+A_{12} \,\varepsilon\left(p_1,p_2,\epsilon_2,\epsilon_3\right)\right)=\tilde{\mathcal{X}}_{8}$\\
$A_{21}^2 B_{13} \,\varepsilon\left(p_1,p_2,\epsilon_1,\epsilon_3\right)-A_{12}^2 B_{23} \,\varepsilon\left(p_1,p_2,\epsilon_2,\epsilon_3\right)=\tilde{\mathcal{X}}_{9}$\\

$A_{12} A_{21} \left(B_{23} \,\varepsilon\left(p_1,p_2,\epsilon_1,\epsilon_3\right)-B_{13} \,\varepsilon\left(p_1,p_2,\epsilon_2,\epsilon_3\right)\right)=\tilde{\mathcal{X}}_{10}$\\

$A_{12} A_{21} A_{31} \,\left(\varepsilon\left(p_1,\epsilon_1,\epsilon_2,\epsilon_3\right)+ \,\varepsilon\left(p_2,\epsilon_1,\epsilon_2,\epsilon_3\right)\right)=\tilde{\mathcal{X}}_{11}$\\
$A_{12} A_{21} A_{31}^2 \,\varepsilon\left(p_1,p_2, \epsilon_1,\epsilon_2\right)=\tilde{\mathcal{X}}_{12}$\\
$A_{21} A_{31} A_{12} \, \left(A_{21} \,\varepsilon\left(p_1,p_2,\epsilon_1,\epsilon_3\right)+ A_{12}
   \,\varepsilon\left(p_1,p_2,\epsilon_2,\epsilon_3\right)\right)=\tilde{\mathcal{X}}_{13}$
	\end{tabular}
\caption{All possible parity-odd on-shell three-point amplitudes for two identical, one different  massive spin-2 particles.} \label{MM2M2odd}
\end{table}

Dimensional-dependent relations in $d=4$ -see \cite{Bonifacio:2018vzv} for the details- impose 
\begin{subequations}
\label{eq:odddd}
\begin{align}
 2 \left(m_k^4-4 m^2 m_k^2\right)\tilde{\mathcal{X}}_{1}+\left( 4m^2
   m_k^2-m_k^4\right)&\tilde{\mathcal{X}}_{2}\nonumber\\  -\left(m_k^2-2
   m^2\right){}^2\tilde{\mathcal{X}}_{3}+\left(2m^2
   m_k^2-4m^4\right)&\tilde{\mathcal{X}}_{4}\nonumber\\ -4m^4\tilde{\mathcal{X}}_{5}+4\left(m^2-m_k^2\right)\tilde{\mathcal{X}}_{6}+4m^2\tilde{\mathcal{X}}_{7}&=0\, ,\\  
   \tilde{\mathcal{X}}_{2}-\tilde{\mathcal{X}}_{3}&=0\, ,\\   
      \tilde{\mathcal{X}}_{10}-\tilde{\mathcal{X}}_{11}&=0\, ,\\ 
  m_k^2\tilde{\mathcal{X}}_{7}-m_k^2\tilde{\mathcal{X}}_{8}-2m^2\tilde{\mathcal{X}}_{11}+2\tilde{\mathcal{X}}_{12}&=0\, ,
\end{align}
\end{subequations}
which we can use to ignore four of the $\tilde{\mathcal{X}}_{i}$ involved in \eqref{eq:odddd}, writing them as linear combination of the others.

\subsection{Lagrangian basis}
\label{sec:lagrangian}

To write a Lagrangian basis we recall the expression for the linearized version of the Riemann tensor $R_{\alpha \beta \mu \nu}$ for a spin-2 field $h_{\mu\nu}$ 
\begin{align}
R_{\alpha \beta \mu \nu}=\frac{1}{2}\left[\partial_\mu \partial_\beta h_{\nu \alpha}+\partial_\nu \partial_\alpha h_{\beta \mu}-\partial_\nu \partial_\beta h_{\mu \alpha}-\partial_\mu \partial_\alpha h_{\beta \nu}\right]\, , 
\end{align}
and  define the tensor $F_{\alpha\beta\mu}$  as 
\begin{equation}
F_{\alpha\beta\mu}\equiv \partial_\alpha h_{\beta\mu}-\partial_\beta h_{\alpha\mu}\, .
\end{equation}
Using these two quantities, a Lagrangian basis for the   parity-even on-shell three-point amplitudes introduced in table \ref{MM2M2} is given in table \ref{lagrangianbasis} below.
\begin{table}[H]
\centering
\def\arraystretch{1.5}%  1 is the default, change whatever you need
\begin{tabular}{ c}
Lagrangian basis $\mathcal{L}_{2,2,2}\left(m,m,m_k\right)$    \\ \hline
$h^\mu_{1\nu} h^\nu_{2\alpha} h^\alpha_{3\mu}=\mathcal{L}_1$\\
$F^{1\mu\alpha}_{\beta}F^2_{\nu\mu\alpha}h^{3\beta\nu}=\mathcal{L}_2$\\
$F^{1\mu\alpha}_{\beta}h^{2\beta\nu}F^3_{\nu\mu\alpha}+h^{1\beta\nu}F^{2\mu\alpha}_{\beta}F^3_{\nu\mu\alpha}=\mathcal{L}_3$\\
$ h^{1\mu\alpha}h^{2\nu\beta}R^3_{\mu\nu\alpha\beta}=\mathcal{L}_4$\\
$h^{1\mu\alpha}R^2_{\mu\nu\alpha\beta}h^{3\nu\beta}+R^1_{\mu\nu\alpha\beta}h^{2\nu\beta}h^{3\mu\alpha}=\mathcal{L}_5$\\
$ R^{1\mu\nu\alpha\beta}F^{2\delta}_{\mu\nu}F^3_{\alpha\beta\delta}+F^{1\delta}_{\mu\nu}R^{2\mu\nu\alpha\beta}F^3_{\alpha\beta\delta}=\mathcal{L}_6$\\
$F^1_{\alpha\beta\delta}F^{2\delta}_{\mu\nu}R^{3\mu\nu\alpha\beta}=\mathcal{L}_7$\\
$R^{1\mu\nu}_{\alpha\beta}R^{2\gamma\delta}_{\mu\nu}R^{3\alpha\beta}_{\gamma\delta}=\mathcal{L}_{8}$\\
	\end{tabular}
\caption{Lagrangians describing parity-even three-point amplitudes for two identical, one different massive spin-2 particles} \label{lagrangianbasis}
\end{table}
The relation with table \ref{MM2M2}  is given by
\begin{subequations}
\label{eq:rellagon}
\begin{align}
\mathcal{X}_1&=\mathcal{L}_1\, ,\\
\mathcal{X}_2&=\left(m^2-\frac{3}{2} m_k^2\right)\mathcal{L}_1-\mathcal{L}_2+\mathcal{L}_3-\mathcal{L}_4\, ,\\
\mathcal{X}_3&=\left(m_k^2-2m^2\right)\mathcal{L}_1 +2 \mathcal{L}_2-\mathcal{L}_5\, ,\\
2\mathcal{X}_4&=\left(2m^2-m_k^2\right)\mathcal{L}_1-2 \mathcal{L}_2-\mathcal{L}_4+\mathcal{L}_5\, ,\\
\mathcal{X}_5&= m_k^2\mathcal{L}_1-\mathcal{L}_3+\mathcal{L}_4\, ,\\
4\mathcal{X}_6&=3m_k^4 \mathcal{L}_1 -2 m_k^2 (\mathcal{L}_3-\mathcal{L}_4)+\mathcal{L}_7\, ,\\
4\mathcal{X}_7&=6m_k^2\left(2 m^2 - m_k^2\right)\mathcal{L}_1-4 m_k^2\mathcal{L}_2\nonumber\\&+2\left( m_k^2-2 m^2\right)\mathcal{L}_3+4 \left(m^2-m_k^2\right)\mathcal{L}_4\nonumber\\&+2 m_k^2\mathcal{L}_5+\mathcal{L}_6\, ,\\
8\mathcal{X}_8&=5m_k^4\left(2 m^2 - m_k^2\right)\mathcal{L}_1-2 m_k^4\mathcal{L}_2\nonumber\\&+2m_k^2\left( m_k^2-2 m^2 \right)\mathcal{L}_3+m_k^2\left(4 m^2 -3 m_k^2\right)\mathcal{L}_4\nonumber\\&+m_k^4\mathcal{L}_5+m_k^2\mathcal{L}_6+\left(2 m^2-m_k^2\right)\mathcal{L}_7-\mathcal{L}_8\, .
\end{align}
\end{subequations}
It is important to bear in mind  that Lagrangians are off-shell quantities. Under field redefinitions or integration by parts they give rise to the same (on-shell) dynamics. This being said, notice that the l.h.s of equation  \eqref{eq:rellagon} is defined on-shell. Therefore, the equality only makes sense when the r.h.s -the linear combination of Lagrangians- is  also evaluated on-shell.

\section{Couplings from GR}
\subsection{Cubic couplings}
\label{app:cubic}

In this appendix we write the three-point interactions that result from plugging the decomposition \eqref{eq:expansion} into the Einstein-Hilbert action. These results were published initially in \cite{Hinterbichler:2013kwa, Bonifacio:2019ioc}. 

To start with, we need to fix the notation. A particle $i$ has momentum $p_i$ and mass $m_i$. We denote its polarisation tensor by $\epsilon_{i}$. This tensor is symmetric and traceless. Formally, when constructing the interactions, we will write the polarisation matrices as a product of vectors $\epsilon_{i}=\epsilon_{i\mu\nu}\equiv \epsilon_{i\mu}\epsilon_{i\nu}$. This does not mean that 
the polarisations matrices have rank one, it is only a trick to keep track of the contractions more easily. To simplify the expressions, we call $A_{ij}\equiv \epsilon_i\cdot p_j$, $B_{ij}=\epsilon_i\cdot\epsilon_j$ and $p_{ij}=p_i\cdot p_j$. Finally, we put the  external legs of the amplitude (the massive spin-2 particles) on-shell, whereas we keep the exchanged particle off-shell.

%Finally, we denote by $\mathcal{M}_{a,b,c} (m_a,m_b,m_c)$ a three-point interaction of particles with spin $a$, $b$, $c$ and masses $m_a$, $m_b$, $m_c$, where the particle $c$ will be the one exchanged.

This being said, the vertices involving (at least) two massive spin-2  are:
\begin{itemize}
\item Three massive spin-2 particles $\{h^{n_1}, h^{n_2}, h^{n_3}\}$
\begin{align}
&\mathcal{A}\left( h^{n_1}, h^{n_2}, h^{n_3}\right)=\nonumber\\
&-\frac{g_{n_1n_2n_3}}{4M_D^{\frac{D-2}{2}}}B_{23}\left[4A_{13}\left(B_{23}A_{12}-2B_{12}A_{32}\right)\right.\nonumber\\&\left.+(2p_{12}-m_{n_1}^2)B_{12}B_{23}\right]+\rm{5\,  permutations}\, ,
\end{align}
where we have implicitly assigned the numbers
\begin{align}
\{1,2,3\}\equiv \{h^{n_1}, h^{n_2}, h^{n_3}\}\, ,
\end{align}  
-we will also do this for the other interactions- and have introduced the triple overlap integrals
\begin{align}
g_{n_1n_2n_3}=\int_{ X_{D-4}}\psi_{n_1}\psi_{n_2}\psi_{n_3}\, d\text{vol}_{X_{D-4}}\, .
\end{align}
\item Two massive spin-2 particles and the graviton $\{h^{n_1}, h^{n_2}, h^{0}\}$
\begin{align}
\label{eq:hhho}
&\mathcal{A}\left( h^{n_1}, h^{n_2}, h^{0}\right)=\nonumber\\
&\frac{1}{4M_d^{\frac{d-2}{2}}}\delta_{n_1,n_2}\left[8\left(B_{23}A_{12}-B_{13}A_{21}\right)\left(B_{23}A_{12}+B_{12}A_{31}\right)\right.\nonumber\\ &\left.+\left(B_{12}\right)^2\left(p_{33}B_{33}-4A_{31}A_{32}\right)\right]+1\leftrightarrow2\, .
\end{align}
\item Two massive spin-2 particles and one spin-1 particle  $\{h^{n_1}, h^{n_2}, A^{i_3}_\mu\}$
\begin{align}
&\mathcal{A}\left( h^{n_1}, h^{n_2}, A^{i_3}_\mu\right)=\nonumber\\
\label{eq:hha}
&\frac{\sqrt{2}}{M_D^{\frac{D-2}{2}}}g_{n_1n_2i_3}\left(2B_{12}B_{13}A_{21}-B_{12}^2A_{31}\right.\nonumber \\ &\left.-\left(1\leftrightarrow2\right)\right)\, ,
\end{align}
with
\begin{align}
g_{n_1n_2i_3}=\int_{ X_{D-4}}\partial^a \psi_{n_1}\psi_{n_2}Y_{a,i_3}\, d\text{vol}_{X_{D-4}}\, ,
\end{align}
being antisymmetric in the first two indices -so necessarily $n_1\neq n_2$-. 
\item Two massive spin-2 particles and one scalar particle $\{h^{n_1}, h^{n_2},\phi\}$
\begin{align}
&\mathcal{A}\left( h^{n_1}, h^{n_2}, \phi\right)
\propto \phi B_{12}^2\, .
\end{align}
As commented  in \cite{Kundu:2023cof},  for any polarization of the spin-2 particles this term scales with $s$ as $s^n$, $n\leq 2$: it does not contribute  to the CRG equations. For this reason, for our purposes it is enough  to write the part that gives the dependence on the kinematics.
\end{itemize}
\subsection{Quartic  couplings}
\label{app:contact}

Following the notation introduced in the previous section, the on-shell 4-point interaction between any four massive spin-2 particles $\{h^{n_1}, h^{n_2}, h^{n_3}, h^{n_4}\}$ that are part of the KK tower of the graviton is given by 
\begin{align}
\label{eq:contact}
&\mathcal{A}_{\text{contact}}\left( h^{n_1}, h^{n_2}, h^{n_3}, h^{n_4}\right)=\nonumber\\
&\frac{1}{M_D^{D-2}}\left[g_{n_1n_2n_3n_4}B_{12}B_{34}\left(\left(p_{12}-m_{n_1}^2\right)B_{14}B_{23}\right.\right.\nonumber \\ &\left.+A_{14}\left[B_{34}A_{23}-B_{24}A_{32}-B_{23}\left(A_{41}+2A_{43}\right)\right]\left.\right)\right.\nonumber \\ &\left.+\sum_i\frac{1}{2}m_{n_i}^2g_{n_1n_2n_i}g_{n_3n_4n_i}B_{12}B_{34}B_{14}B_{23}\right]\nonumber\\+&\rm{23\, permutations}\, ,
\end{align}
where
\begin{align}
g_{n_1n_2n_3n_4}=\int_{ X_{D-4}}\psi_{n_1}\psi_{n_2}\psi_{n_3}\psi_{n_4}\, d\text{vol}_{X_{D-4}}\, .
\end{align}

\section{Kinematics}
\label{app:kinematics}

In this appendix we will write the definitions of the  variables used to compute the $2\rightarrow 2$ scattering of section \ref{subsec:resuls}. 

The incoming particles are labelled by $1$ and $2$, and outgoing particles by $3$ and $4$. Momentum conservation requires
\begin{align}
p_1+p_2=p_3+p_4\, ,
\end{align}
where we take
\begin{align}
p_i^\mu=\left(E_i,p_i \sin\theta_i,0,p_i\cos\theta_i\right)\, ,
\end{align}
with $E_i^2=p_i^2+m_i^2$ and $\theta_1=0$, $\theta_2=\pi$, $\theta_3=\theta$, $\theta_4=\theta-\pi$. The Mandelstam variables are
\begin{align}
\label{mandelstam}
s=&-\left(p_1+p_2\right)^2\, ,	&	t&=-\left(p_1-p_3\right)^2\, ,	&	u&=-\left(p_1-p_4\right)^2\, ,
\end{align}
and we are taking the metric $\eta=\text{diag}\left(-1,1,1,1\right)$. They satisfy
\begin{align}
s+t+u=m_1^2+m_2^2+m_3^2+m_4^2\, ,
\end{align}
with $m_i$ the mass of the  particle $i$. When  $m_1=m_2=m_3=m_4=m$ -and $E_i=E\, , p_i=p$- the Mandelstam variables have the simple expressions
\begin{align}
s&=4E^2\, ,	&	cos\theta=1-\frac{2t}{4m^2-s}\, .
\end{align}
To construct the polarisation matrices of the spin-2 particles, we first introduce the polarisation vectors:
\begin{subequations}
\begin{align}
\epsilon_{1}^\mu(p_i)&=\left(0, 0, 1, 0\right)\\
\epsilon_{2}^\mu(p_i)&=\left(0, \cos\left(\theta_i\right),0, -\sin\left(\theta_i\right)\right)\\
\epsilon_{3}^\mu(p_i)&=\frac{1}{m_i}\left(p_i,e_i \sin\left(\theta_i\right),0, e_i\sin\left(\theta_i\right)\right)\, ,
\end{align}
\end{subequations}
from which the polarisation matrices can be constructed, adapting the conventions of \cite{Bonifacio:2017nnt},  as 
\begin{subequations}
\label{eq:pol2def}
    \begin{align}
\epsilon^{\mu\nu}_{T_,1}(p_i)&=\frac{1}{\sqrt{2}}\left(\epsilon_{1}^\mu(p_i)\epsilon_{1}^\nu(p_i)-\epsilon_{2}^\mu(p_i)\epsilon_{2}^\nu(p_i)\right)\, , \\
\epsilon^{\mu\nu}_{T_,2}(p_i)&=\frac{1}{\sqrt{2}}\left(\epsilon_{1}^\mu(p_i)\epsilon_{2}^\nu(p_i)+\epsilon_{2}^\mu(p_i)\epsilon_{1}^\nu(p_i)\right)\, , \\
\epsilon^{\mu\nu}_{V_,1}(p_i)&=\frac{1}{\sqrt{2}}\left(\epsilon_{3}^\mu(p_i)\epsilon_{1}^\nu(p_i)+\epsilon_{1}^\mu(p_i)\epsilon_{3}^\nu(p_i)\right) \, , \\
\epsilon^{\mu\nu}_{V_,2}(p_i)&=\frac{1}{\sqrt{2}}\left(\epsilon_{3}^\mu(p_i)\epsilon_{2}^\nu(p_i)+\epsilon_{2}^\mu(p_i)\epsilon_{3}^\nu(p_i)\right)\, , \\
\epsilon^{\mu\nu}_{S}(p_i)&=\sqrt{\frac{3}{2}} \left(\epsilon_{3}^\mu(p_i)\epsilon_{3}^\nu(p_i)-\frac{1}{3}\left(\eta^{\mu\nu}+\frac{1}{m_i^2}p_i^\mu p_i^\nu\right)\right)\, ,
\end{align}
\end{subequations}
where $T$, $V$ and $S$ stand for tensor, vector and scalar polarizations, respectively.

Regarding the propagators, the propagator of a scalar particle with mass $M$ is
\begin{align}
    \frac{-i}{p^2+M^2-i\epsilon}\, .
\end{align}
For a massive spin-1 particle, first we need to introduce the projector
\begin{align}
    \Pi_{\mu\nu}\left(\tilde m\right)=\eta_{\mu\nu}+\frac{p_\mu p_\nu}{\tilde m^2}\, ,
\end{align}
from which one can write the propagator of a  massive spin-1 particle with mass $m_1$ as
\begin{align}
  P_{\mu\nu}= \frac{ -i\Pi_{\mu\nu}\left( m_1\right)}{p^2+m_1^2-i\epsilon}\, .
\end{align}
Finally, the propagator of a massive spin-2 particle of mass $m$ is
\begin{align}
     P_{\mu_1\mu_2,\nu_1\nu_2}=\frac{-i}{2}  \frac{\Pi_{\mu_1\nu_1}\Pi_{\mu_2\nu_2}+\Pi_{\mu_1\nu_2}\Pi_{\mu_2\nu_1}-\frac{2}{3}\Pi_{\mu_1\mu_2}\Pi_{\nu_1\nu_2}}{p^2+m^2-i\epsilon}\, ,
\end{align}
with $\Pi_{\nu_1\nu_2}=\Pi_{\nu_1\nu_2}\left( m\right)$, whereas for a massless spin-2 (in de Donder gauge) it reads 
\begin{align}
     \tilde P_{\mu_1\mu_2,\nu_1\nu_2}=\frac{-i}{2}  \frac{\eta_{\mu_1\nu_1}\eta_{\mu_2\nu_2}+\eta_{\mu_1\nu_2}\eta_{\mu_2\nu_1}-\eta_{\mu_1\mu_2}\eta_{\nu_1\nu_2}}{p^2-i\epsilon}\, .
\end{align}

\bibliography{papers}

\end{document}